\documentclass[12pt]{iopart}
\usepackage{iopams}
\usepackage{amsfonts}
\usepackage{graphicx}
\usepackage{epsfig}
\usepackage{graphics}
\makeatletter
\newcommand{\row}[1]%
{\mathord{\buildrel{\lower3pt%
\hbox{$\scriptscriptstyle\rightarrow$}}\over #1}}
\newcommand{\col}[1]{{#1^{\raisebox{2pt}[\height]%
{$\scriptstyle\downarrow$}}}}
\newcommand{\dyadic}[1]{\mathord{\dyadic@rrow{#1}}}
\newcommand{\dyadic@rrow}[1]{
\begin{picture}(12,12)(-1,0)
\put(-2,10){\makebox(0,0)[t]{$\scriptscriptstyle\downarrow$}}
\put(-2,11){\makebox(0,0)[l]{$\scriptscriptstyle\longrightarrow$}}
\put(5,0){\makebox(0,0)[b]{$#1$}}
\end{picture}
}

\newcommand{\bra}[1]{\bigl\langle #1 \bigr|}
\newcommand{\ket}[1]{\bigl| #1 \bigr\rangle}
\newcommand{\expect}[1]{\left\langle #1 \right\rangle}
\newcommand{\trans}[1]{{#1}^{\raisebox{-1pt}{\scriptsize\textrm{T}}}}
\newcommand{\PRA}[3]{Phys.\ Rev.\ A  \textbf{#1}, #2 (#3)}
\newcommand{\plA}[3]{Phys.\ Lett.\ \textbf{A#1}, #2 (#3)}
\newcommand{\JMO}[3]{J. Mod.\ Opt.\ \textbf{#1}, #2 (#3)}

\makeatletter

\begin{document}

\title{Abruption of entanglement and quantum communication
through noise channels}
\author{ Nasser Metwally }
\address
{ Math. Dept., Faculty of Science, South Valley University, Aswan,
Egypt.}
 \ead{NMetwally@gmail.com}

 \begin{abstract}
We investigate the dynamics of  two qubits state through the
Bloch channel.
  Starting from partially entangled states as
input state,  the output states are more robust compared with
those obtained from initial  maximally entangled states. Also the
survivability of  entanglement increased as the absolute
equilibrium values of the channel increased or the ratio between
the longitudinal and transverse relaxation times gets smaller. The
ability of using the output states as quantum channels to perform
quantum teleportation is investigated. The   useful output states
are used to send information between two users by using the
original  quantum
teleportation protocol.\\

\hspace{1cm} 
\end{abstract}
\maketitle
\section{Introduction}

  Nowadays information can be  stored, transmitted  and manipulated  by
qubits. The most important kinds of qubits are the entangled ones.
Although it is  possible to generate useful entangled states  for
 quantum information purposes,  decoherence processes result in
 shortening their survival. This, in turn affects efficiency of performing such  tasks as in quantum teleportation
\cite{Briegel,HNR}. So, finding a robust scheme for quantum
information tasks is very important \cite{Luo}.

Decoherence  represents  an inevitable process which  causes
 entanglement to be  fragile. There is a new kind of decay called the death of entanglement resulting
from classical noise has been discussed recently
\cite{Eberly1,Aolita,Pineda}.  In reality there are several ways
causing the indescribable decoherence. For  example, the
interaction of qubits with its surroundings \cite{Ekert}, device
imperfections \cite{Gero,Sug}, the decay due to  spontaneous,
emission and the noisy channel \cite{MFS}.

So, investigating the dynamics of entanglement in the presence  of
decoherence is one of the most important tasks in quantum
computation and information. In the present work,  we  examine
 some intrinsic properties of the dynamics of a two qubit
state passing through Bloch channels. The decoherence of
entanglement  and information in these types
 of channel has been investigated \cite{Ban-1, Ban-2}, where,
the case of only one qubit passing through the Bloch channel is
studied. In our contribution, we assume that there is a source
that  supplies us with a two qubit state. One qubit is sent to the
user, Alice and the second qubit is sent to  Bob. Then the two
qubits are forced to  be sent through the Bloch channel. Our study
focus on the  properties of the output state  from different
directions as we shall see later.

 The paper is organized
as follows: In Sec.$2$, the evolution  of a general two-qubit
state passes through Bloch channel is examined analytically.
 Sec.$3$ is
devoted for numerical calculations, where
 the survival degree of entanglement is
quantified and  the phenomena of the decay and sudden death of
entanglement are examined.
 Use the output state
for quantum teleportation is studied in Sec.$4$.  Finally, a
conclusion is given  in Sec$.5$.

\section{The Model}
 The characterization
of the 2-qubit states produced by some source requires
experimental determination of 15 real parameters. Each qubit is
determined by $3$ parameters, representing the Bloch vectors, and
the other $9$ parameters represent the correlation tensor. Analogs
of Pauli's spin operators are used for the description of the
individual qubits; the set $\sigma_1,\sigma_2,\sigma_3$ for
Alice's qubit and $\tau_1,\tau_2, \tau_3$ for Bob's qubit. Any two
qubit state is  described by \cite{Gunter,NM, wang},

\begin{equation}
\rho_{ab}=\frac{1}{4}(1+\row{A}\cdot\col{\sigma}+\row{B}\cdot\col\tau+\row\sigma
\cdot\dyadic{C}\cdot\col\tau),
\end{equation}
where $\row\sigma$ and $\row\tau$ are the Pauli's spin vector of
the first and the second qubits respectively. The statistical
operator for the individual qubit are specified by their Bloch
vectors, $\row{A}=\expect{\row\sigma}$ and
$\row{B}=\expect{\row\tau}$. The cross dyadic $\dyadic{C}$ is
represented  by a $3\times 3$ matrix. it describes the correlation
between the first qubit,
$\rho_a=tr_{b}\{\rho_{ab}\}=\frac{1}{2}(1+\row{A}\cdot\col\sigma)$
and the second qubit
$\rho_b=tr_{a}\{\rho_{ab}\}=\frac{1}{2}(1+\row{B}\cdot\col\tau)$.
The Bloch vectors and the cross dyadic are given by
\begin{eqnarray}
\row{A}&=&(A_1,A_2,A_3) ,\quad \row{B}=(B_1,B_2,B_3),\quad
\mbox{and}~ \dyadic{C}= \left(
\begin{array}{ccc}
c_{11}&c_{12}&c_{13}\\
c_{21}&c_{22}&c_{23}\\
c_{31}&c_{32}&c_{33}
\end{array}
\right)
\end{eqnarray}

 Let us consider that each qubit is forced to pass in a
Bloch channel. These channels are defined by the Bloch equations
\cite{Ban-1}, for the first qubit,
\begin{eqnarray}\label{ch1}
\frac{d}{dt}\expect{\sigma_1}_t&=&-\frac{1}{T_{2a}}\expect{\sigma_1}_t,~
\frac{d}{dt}\expect{\sigma_2}_t=-\frac{1}{T_{2a}}\expect{\sigma_2}_t,
\nonumber\\
\frac{d}{dt}\expect{\sigma_3}_t&=&-\frac{1}{T_{1a}}(\expect{\sigma_3}_t-\expect{\sigma_3}_{eq}),
\end{eqnarray}
while for the second qubit, they are given by
\begin{eqnarray}\label{ch2}
\frac{d}{dt}\expect{\tau_1}_t&=&-\frac{1}{T_{2b}}\expect{\tau_1}_t,~
\frac{d}{dt}\expect{\tau_2}_t=-\frac{1}{T_{2b}}\expect{\tau_2}_t,
\nonumber\\
\frac{d}{dt}\expect{\tau_3}_t&=&-\frac{1}{T_{1b}}(\expect{\tau_3}_t-\expect{\tau_3}_{eq}),
\end{eqnarray}
where $T_{1i}$ and $T_{2i}$,~ $i=a,b$  are the longitudinal and
transverse relaxation times for Alice and Bob's qubit, and
$\expect{\sigma_3}_{eq}$,~ $\expect{\tau_3}_{eq}$ are the
equilibrium values of $\expect{\sigma_3}_t$ and
$\expect{\tau_3}_t$ respectively. Now, we  assume that Alice's
qubit $\rho_a$ and Bob's qubit $\rho_b$ pass in the channels
(\ref{ch1}), and (\ref{ch2}) respectively. Then the output state
of the two qubit is defined by their new Bloch vectors
\begin{eqnarray}\label{out1}
\tilde{\row{A}}(t)&=&(A_1\beta_1,~-\beta_1 A_2,~\gamma_1
A_3+(1-\gamma_1)\expect{\sigma_3}_{eq}),
\nonumber\\
\tilde{\row{B}}(t)&=&(\beta_2 B_1,~-\beta_2 B_2,~\gamma_2
B_3+(1-\gamma_2)\expect{\tau_3}_{eq}).
\end{eqnarray}
In addition we present new correlation tensor
\begin{eqnarray}\label{out2}
\tilde C_{11}(t)&=&\beta_1\beta_2 C_{11},\quad \tilde
C_{12}(t)=-C_{12}\beta_1\beta_2,
 \nonumber\\
\tilde
C_{13}(t)&=&\beta_1\gamma_2C_{13}+\beta_1(1-\gamma_2)\expect{\sigma_3}_{eq}A_1,
\nonumber\\
\tilde C_{21}(t)&=&-C_{12}\beta_1\beta_2,~
 \tilde C_{22}(t)=C_{22}\beta_1\beta_2,
 \nonumber\\
 \tilde C_{23}(t)&=&-\beta_1\gamma_2C_{23}-\beta_1(1-\gamma_2)\expect{\tau_3}_{eq}A_2,
\nonumber\\
 \tilde C_{31}(t)&=&\beta_2\gamma_1C_{31}+\beta_2(1-\gamma_1)\expect{\sigma_3}_{eq}B_1,
 \nonumber\\
\tilde
 C_{32}(t)&=&-\beta_2\gamma_1C_{32}-\beta_2(1-\gamma_1)\expect{\sigma_3}_{eq}B_2,
\nonumber\\
 \tilde C_{33}(t)&=&\gamma_1\gamma_2C_{33}+(1-\gamma_1)(1-\gamma_2)\expect{\sigma_3}_{eq}\expect{\tau_3}_{eq}
\nonumber\\
&&+\gamma_1(1-\gamma_2)\expect{\tau_3}_{eq} A_3
+\gamma_2(1-\gamma_1)\expect{\sigma_3}_{eq}B_3,
\end{eqnarray}
where $\gamma_i=exp\{-\frac{t}{T_{1i}}\}, ~
\beta_i=exp\{-\frac{t}{T_{2i}}\},~\mbox{and} ~i=a,b$.

Before  starting  our investigation, it is important to shed some
light on the positivity of the Bloch channel. A quantum channel,
$\mathcal{B}_t$ has the positivity property  if
$(i)~\mathcal{B}_t\{\rho\}$ is positive,
$(ii)~tr{\mathcal{B}_t}\{\rho\}=tr\{\rho\}$ and
$(ii)~\mathcal{B}_t\otimes I_N$ is positive. The latter property
guarantees that  the channel $\mathcal{B}_t$ is completely
positive, see for example \cite{BFR}. The conditions $(i)$, and
$(ii)$ are satisfied directly for the Bloch channel, while  for
the third criteria, the channel is completely positive if the
following inequalities are satisfied for each  qubit,
\begin{eqnarray}
\gamma_1>\frac{2\beta_1+\expect{\sigma_3}}{\expect{\sigma_3}+2},\quad
\beta_1<\frac{\sqrt{1-\expect{\sigma_3}}}{4}(1-\gamma_1), ~\quad
\nonumber\\
\gamma_2>\frac{2\beta_2+\expect{\tau_3}}{\expect{\tau_3}+2},\quad
\beta_2<\frac{\sqrt{1-\expect{\tau_3}}}{4}(1-\gamma_2). ~\quad
\end{eqnarray}

Now, we   investigate some  properties of  the output state by
considering  a class of  maximally entangled states and partially
entangled states. These two classes can be driven from a class of
a generic pure two qubit states.
\section{A generic two-qubit pure state}
 The generic two qubit pure state,$\rho_{p}$ is defined
by,
\begin{eqnarray}\label{pure1}
\row A&=&(0,0,p),\quad \row B=(0,0,-p), \nonumber\\
C_{11}&=&-q,~C_{12}=C_{13}=0, \nonumber\\
  C_{21}&=&0,~C_{22}=-q,~C_{23}=0, \nonumber\\
C_{31}&=&0~,C_{32}=0,~C_{33}=-1,
\end{eqnarray}
where $0<p<1$ and $q=\sqrt{1-p^2}$. This class of states
represents the Bell states for $p=0$ and $q=1$ and a product state
for $p=1$ and $q=0$. The degree of  entanglement for this class is
given by its concurrence\cite{Nasser}.
 By using the initial Bloch vectors  $\row{A}$ and $\row{B}$ in  equation
(\ref{out1}), one gets the new Bloch vectors for the output state
as,
\begin{eqnarray}\label{pure1-1}
\tilde{\row{A}}(t)&=&(0,0,p\gamma_1+(1-\gamma_1)\expect{\sigma_3}_{eq}),\quad
\nonumber\\
 \tilde{\row
B}(t)&=&(0,0,-p\gamma_2+(1-\gamma_2)\expect{\tau_3}_{eq}).
\end{eqnarray}
Similarly, by  using the initial  non zero elements of the
correlation tensor  from (\ref{pure1}) in Eq.(\ref{out2}), one
gets, the new elements of the correlation tensor as,
\begin{eqnarray}\label{pure1-2}
\tilde C_{11}(t)&=&-q\beta_1\beta_2,\quad \tilde
C_{22}(t)=-q\beta_1\beta_2,
 \nonumber\\
\tilde
C_{33}(t)&=&-\gamma_1\gamma_2+(1-\gamma_1)(1-\gamma_2)\expect{\sigma_3}_{eq}\expect{\tau_3}_{eq}
\nonumber\\
&&
+p\left[\gamma_1(1-\gamma_2)\expect{\sigma_3}_{eq}-\gamma_2(1-\gamma_1)\expect{\tau_3}_{eq}\right].
\end{eqnarray}
Now, we study the separability of the $\rho_p^{out}$ which is
defined by its new Bloch vectors (\ref{pure1-1}) and the non-zero
elements of the correlation tensor~(\ref{pure1-2}). To do this, we
apply the partial transpose criterion PPT, where the state is
separable if its partial transpose is nonnegative \cite{Peres}.
The output state, $\rho_p^{out}$ is entangled if it violates the
PPT criterion which is given by

\begin{equation}
\mbox{PPT}=\rho_{11}\rho_{44}-\rho_{23}\rho_{32}>0,
\end{equation}
where
\begin{eqnarray}
\rho_{11}&=&\frac{1}{4}\left[(1-\gamma_1\gamma_2)+(1-\gamma_1)\expect{\sigma_3}_{eq}+(1-\gamma_2)\expect{\tau_3}_{eq}
+\Gamma\right]
 \nonumber\\
&+&\frac{p}{4}\left[(\gamma_1-\gamma_2)+\gamma_1(1-\gamma_2)\expect{\sigma_3}_{eq}-
\gamma_2(1-\gamma_1)\expect{\tau_3}_{eq}\right],
\nonumber\\
\rho_{44}&=&\frac{1}{4}\left[(1-\gamma_1\gamma_2)-(1-\gamma_1)\expect{\sigma_3}_{eq}-(1-\gamma_2)\expect{\tau_3}_{eq}
+\Gamma\right]
 \nonumber\\
&+&\frac{p}{4}\left[-(\gamma_1-\gamma_2)+\gamma_1(1-\gamma_2)\expect{\sigma_3}_{eq}-
\gamma_2(1-\gamma_1)\expect{\tau_3)}_{eq}\right],
\nonumber\\
\rho_{23}&=&\frac{1}{4}\left[(1+\gamma_1\gamma_2)-(1-\gamma_1)\expect{\sigma_3}_{eq}+(1-\gamma_2)\expect{\tau_3}_{eq}
-\Gamma\right]
 \nonumber\\
&+&\frac{p}{4}\left[(\gamma_1+\gamma_2)-\gamma_1(1-\gamma_2)\expect{\sigma_3}_{eq}+
\gamma_2(1-\gamma_1)\expect{\sigma_3}_{eq}\right],
\nonumber\\
\rho_{32}&=&\frac{1}{4}\left[(1+\gamma_1\gamma_2)+(1-\gamma_1)\expect{\sigma_3}_{eq}-(1-\gamma_2)\expect{\tau_3}_{eq}
-\Gamma\right]
 \nonumber\\
&+&\frac{p}{4}\left[-(\gamma_1+\gamma_2)+\gamma_1(1-\gamma_2)\expect{\sigma_3}_{eq}-
\gamma_2(1-\gamma_1)\expect{\tau_3}_{eq}\right],
\end{eqnarray}
$\Gamma=(1-\gamma_1)(1-\gamma_2)\expect{\sigma_3}_{eq}\expect{\tau_3}_{eq}$.

To quantify the  amount of entanglement contained in the entangled
states, we  use   a measure introduced  by  Zyczkowski et. al
\cite{Zyc}. This measure states that if the eigenvalues of the
partial transpose are given by $\lambda_{\ell}, \ell=1,2,3,4$,
then the degree of entanglement, DOE is defined by
\begin{equation}
DOE=\sum_{\ell=1}^{4}{|\lambda_\ell|-1}.
\end{equation}

\subsection{Maximally entangled states}
This class of states is obtained from the  input state
(\ref{pure1}) for $p=1$. In this case the output state is defined
by its Bloch vectors, and the non-zero elements of the correlation
tensor,
\begin{eqnarray}\label{eq:max}
\tilde{\row{A}}(t)&=&(0,0,(1-\gamma_1)\expect{\sigma_3}_{eq}),
\tilde{\row{B}}(t)=(0,0,(1-\gamma_2)\expect{\tau_3}_{eq})~,
\nonumber\\
  \tilde C_{11}(t)&=&\tilde C_{22}(t)=-q\beta_1\beta_2,
\nonumber\\
\tilde
C_{33}(t)&=&\Gamma+\gamma_1(1-\gamma_2)\expect{\sigma_3}_{eq}-\gamma_2(1-\gamma_1)\expect{\tau_3}_{eq}-\gamma_1\gamma_2.
\end{eqnarray}
Now, we examine the separability of the output state
(\ref{eq:max}). To do this, we plot the partial transpose
criterion, PPT for some fixed values of
$\expect{\tau_3}_{eq}=-0.5$ and $\alpha_i=T_{1i}/T_{2i}=2.5$,  and
for different values of the $ \expect{\sigma_3}_{eq}$. Fig.(1)
shows that for small values of $\expect{\sigma_3}_{eq}$, the
output state (\ref{eq:max}) turns into a separable state quickly.
However as the absolute equilibrium values of the first qubit,
$\expect{\sigma_3}_{eq}$ increase the entangled time, the time in
which the state is entangled, increases. In other words, the
output state  (\ref{eq:max}) is more robust  for large values of
 $\expect{\sigma_3}_{eq}$.

\begin{figure}
  \begin{center}
 \includegraphics[width=30pc,height=15pc]{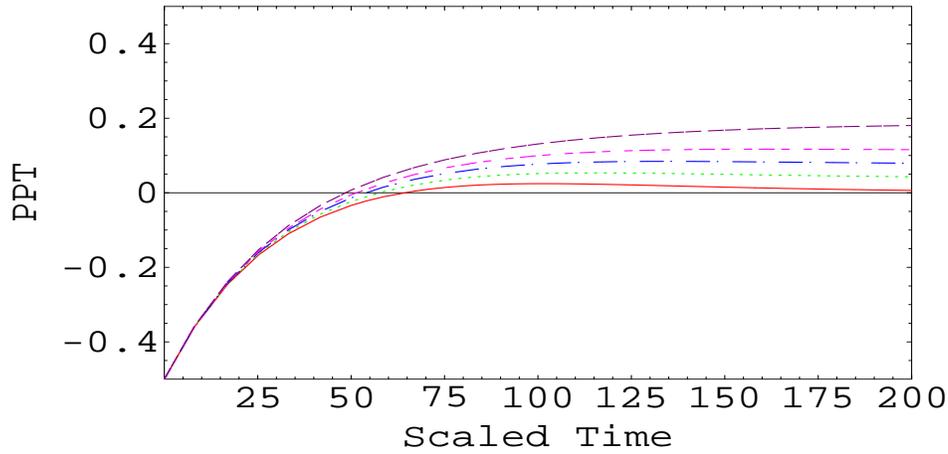}
   \caption{ The PPT criterion for the state (\ref{eq:max}),
   where $p=0,~\expect{\tau_3}_{eq}=-0.5,~ \alpha_i=T_{1i}/T_{2i}=2.5$
   and $\expect{\sigma_3}_{eq}=1,0.9,0.8,0.7,0.5$  for the solid, dot, dashed-dot,
     small-dash and long-dash curves respectively.}
  \end{center}
\end{figure}

\begin{figure}
  \begin{center}
  \includegraphics[width=30pc,height=15pc]{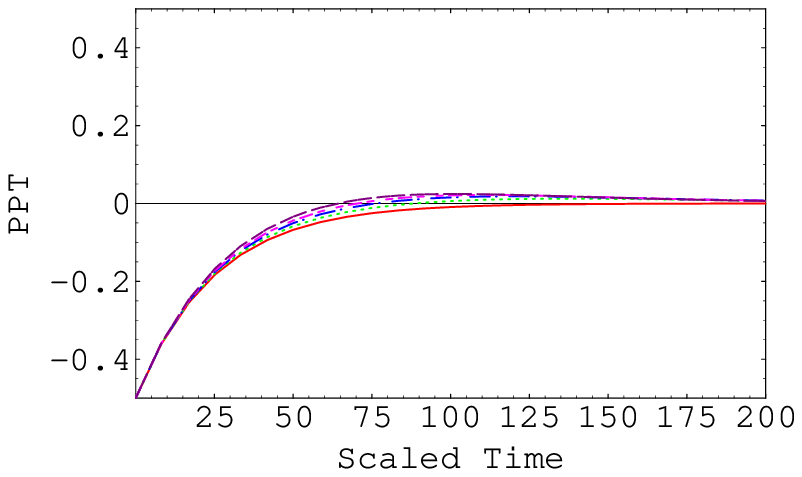}
   \caption{The same as Fig$1$, but for $\expect{\tau_3}_{eq}=1$.}
  \end{center}
\end{figure}

The PPT criterion is examined for large values of the absolute
equilibrium values of the second  qubit, where we consider $
\expect{\tau_3}_{eq}=1$.  In Fig.$2$, the long living entanglement
is observed, where the robustness of the output state
(\ref{eq:max}) is much better than that  depicted in Fig.$1$.
Therefore, by increasing the absolute equilibrium values of the
two qubits, the living time of entanglement  and the robustness of
the output state are increased. Furthermore, the phenomena of
entanglement-breaking,
 where the entangled state  evolves
to a separable state, is observed as one decreases these absolute
equilibrium values.
\begin{figure}
  \begin{center}
 \includegraphics[width=30pc,height=15pc]{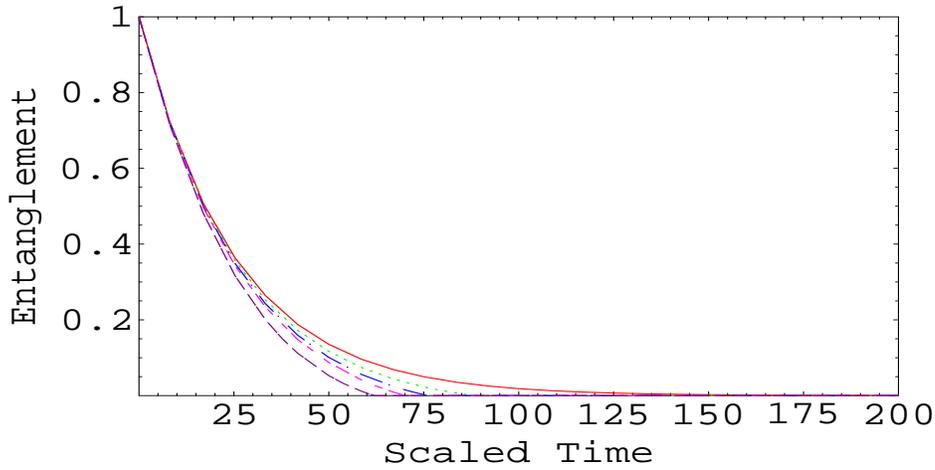}
   \caption{ The degree of entanglement where the parameters are
   the same as Fig$1.$ while $\expect{\tau_3}_{eq}=1$.}
  \end{center}
\end{figure}

\begin{figure}
  \begin{center}
 \includegraphics[width=30pc,height=15pc]{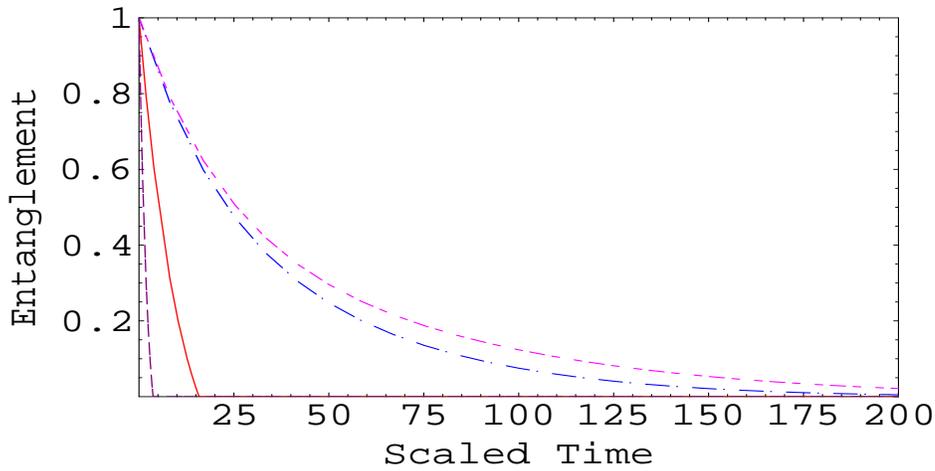}
    \caption{The degree of entanglement for different values of
    the ratio between the longitudinal and transverse time
    $\alpha_i$.
   The parameters are $\expect{\sigma_3}_{eq}=1, \expect{\tau_3}_{eq}=-0.5$ and
   $\alpha_i=2.5,0.5, 0.33,0.25$ for the long-dash, solid,
    dash-dot , and dash-dash curves respectively.}
  \end{center}
\end{figure}

Now, we  investigate the effect of the  equilibrium values on the
survival amount of entanglement for the output state
(\ref{eq:max}).  For the numerical calculations we consider the
case where $\expect{\tau_3}_{eq}=-0.5$, $\alpha_i=\frac{1}{2},~
i=1,2$. Fig.$3$, shows the dynamics of entanglement for different
values $\expect{\sigma_3}_{eq}$. For small values of
$\expect{\sigma_3}_{eq}$, the degree of entanglement decays faster
and the entangled time decreases. On the other hand, the decay of
entanglement is smooth  and the time of living entanglement
increases for larger values of $\expect{\sigma_3}_{eq}$.

The dynamic of entanglement for different values the parameter
$\alpha_i$ is shown in Fig.$4$.
 It is clear that as $\alpha_i$
increases, i.e the longitudinal  relaxation time is larger than
the transverse relaxation time for both qubits, the  entanglement
decays much faster  as compared with in Fig.$3$. Moreover, the
phenomena of the  sudden death of entanglement is observed
\cite{Eberly1,Aolita}.

\subsection{Partially entangled states}
In this section, we consider a class of non-maximally entangled
states. In our calculations we consider a class of partially
entangled states with $p=0.5$. Also, we investigate the dynamic of
PPT criterion and the degree of entanglement, where we use the
same values of the channel parameters.

 Fig.$5$,  shows the effect of the equilibrium
values on the PPT criterion of the output state, where  the
possibility of considering  the Bloch channel as an
entanglement-breaking channel decreases and the time of entangled
increases. Comparing Fig.$1$ with Fig.$5$, we can see that
starting from a partially entangled states the output state is
more robust than starting from a maximally entangled state. This
means that,  the separability and entangled behavior of the input
entangled state, not only depend on the channel parameters but
also on the structure of the input state.
\begin{figure}
  \begin{center}
  \includegraphics[width=30pc,height=15pc]{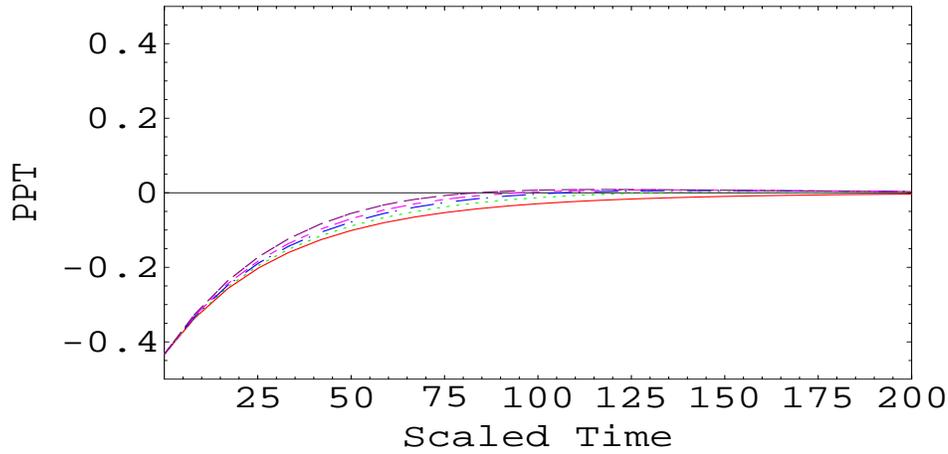}
   \caption{The  same as Fig.$(1)$, but for a partially entangled state.}
  \end{center}
\end{figure}
\begin{figure}
  \begin{center}
 \includegraphics[width=30pc,height=15pc]{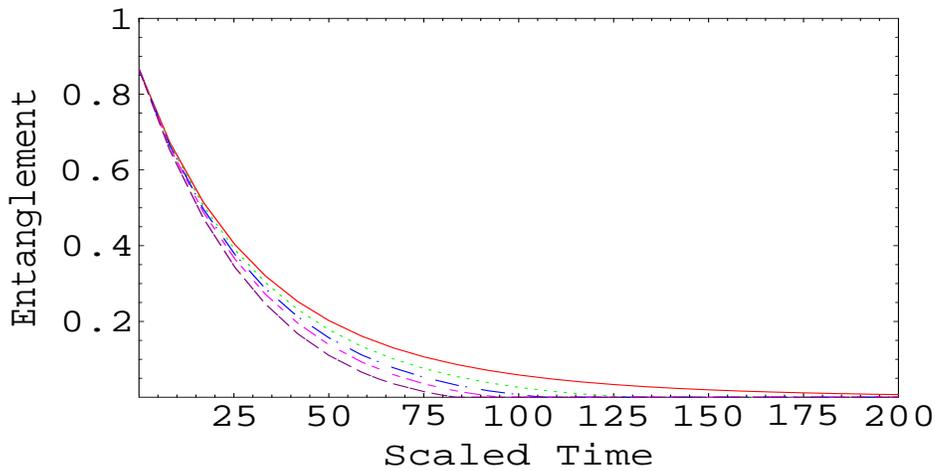}
   \caption{The  same as Fig.$(3)$, but  for a partially entangled state.}
  \end{center}
\end{figure}

 In Fig.$6$, we investigate the dynamics of the entanglement for
 fixed values of
$\expect{\tau_3}_{eq}$ and $\alpha_i$, and for several values of
$\expect{\sigma_3}_{eq}$. The general behavior is the same as that
shown in Fig.$3$, but the entanglement decays more smoothly and
disappears gradually. Therefore the phenomena of sudden death of
entanglement does not show for this class of states. Comparing
Fig.$3$ and Fig.$6$, shows the time of the lived entanglement is
much larger for
 partially entangled state and the entanglement vanishes gradually.
  So  maximally entangled  state is more fragile than
partially entangled state.

\section{Teleportation}

In this section, we examine whether the output state can be used
as a quantum channel to achieve teleportation or not. For this
task, we use  Horodecki's criterion \cite{Horodecki}, where any
mixed spin $\frac{1}{2}$ state is useful for teleportation if
$tr{\sqrt{\trans{\dyadic{C}}\dyadic{C}}}>1$. By using this
criterion, we find that, the output state  which is defined by
(\ref{pure1-1}) and  (\ref{pure1-2}) is available for quantum
teleportation if the following inequality is obeyed

\begin{eqnarray}\label{Telp}
\mbox{Telp}&=&2q^2\beta_1^2\beta_2^2+\bigl[(1-\gamma_1)(1-\gamma_2)\expect{\sigma_3}_{eq}\expect{\tau_3}_{eq}
\nonumber\\
& +&
p\left(\gamma_1(1-\gamma_2)\expect{\tau_3}_{eq}-\gamma_2(1-\gamma_1)\expect{\sigma_3}_{eq}\right)-
\gamma_1\gamma_2\bigr]^2>1.
\end{eqnarray}
\begin{figure}
  \begin{center}
 \includegraphics[width=30pc,height=15pc]{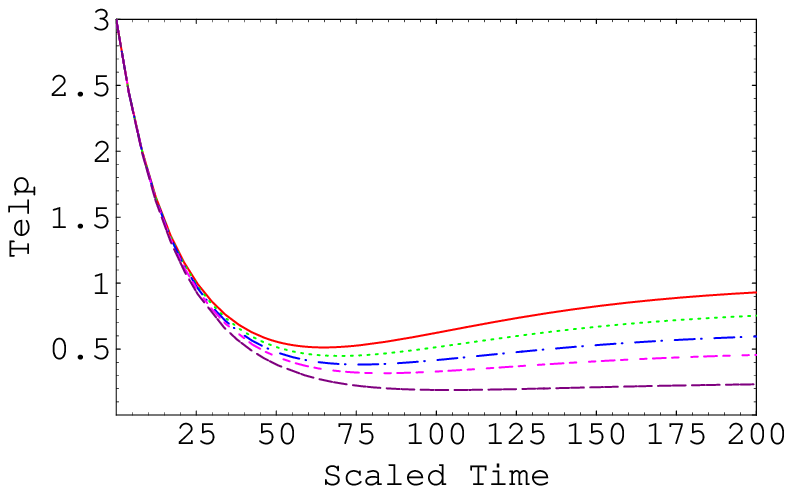}
 \put(-30,160){(a)}\
 \includegraphics[width=30pc,height=15pc]{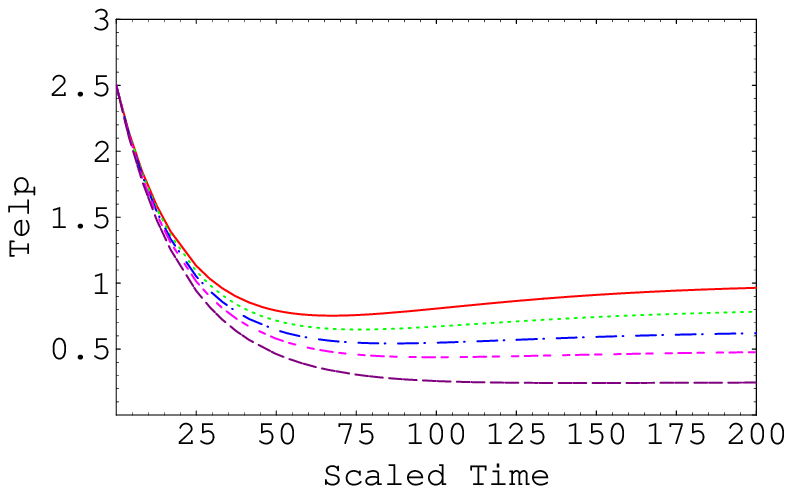}
   \put(-30,160){(b)}
   \caption{The teleportation inequality (\ref{Telp}), where the parameters are the same as Fig.$1$.
    (a)for the maximally entangled state and (b) for the partially entangled state.
 }
  \end{center}
\end{figure}
 Fig.$7a$, shows the behavior of the teleportation inequality (\ref{Telp})
for the output state (\ref{eq:max}). It is clear that  as the
absolute  equilibrium values increase, the possibility of using
this channel for quantum teleportation  increases. As an example
for $ \expect{\sigma_3}_{eq} =1$, the time interval in which the
teleportation  inequality  obeyed  is $[0,25.2]$, while it is
$[0,23.1]$ for $ \expect{\sigma_3}_{eq} =0.5$.

 In Fig.$7b$, we plot the teleportation inequality for the output
 state with $p=0.5$, i.e the partial entangled class.
 In general, the behavior of the teleportation inequality is the same as that
 depicted in Fig.$7b$, but the time to use
 the state as a quantum channel to perform teleportation is larger.
  As an example, for $ \expect{\sigma_3}_{eq} =1$ the teleportation inequality   is
obeyed in the time interval $[0,30.9]$. This is due to that the
output state for a system  prepared initially in a partially
entangled state is more robust than that obtained from a system
initially prepared in a maximum entangled state.

 At this end, we  achieve the quantum teleportation by
 using the output state as a quantum channel.
  Assume that Alice is given an unknown state
 defined by its state vector
 \begin{equation}\label{ste}
 \ket{\Psi}=\lambda_1\ket{0}+\lambda_2\ket{1},
 \end{equation}
where $\lambda^2_1+\lambda^2_2=1$. Now she wants to sent this
state to Bob through their quantum channel. To attain this aim ,
Alice and Bob will use the original teleportation protocol
\cite{bennt}. In this case the total state of the system is
$\rho_{\psi}\otimes\rho^{out}$, where $\psi$ is given by
(\ref{ste}) and $\rho^{out}$ is defined by (\ref{pure1-1}) and
(\ref{pure1-2}). Alice makes a measurement on the given qubit and
her own qubit. Then she sends her results through a classical
channel to Bob. As soon as Bob receives the classical data, he
performs a suitable unitary operation on his qubit and gets the
teleported state. If Alice measures the Bell state,
$\ket{\phi^+}\bra{\phi^+}=\frac{1}{\sqrt{2}}(\ket{00}+\ket{11})$,
then the final state at Bob's hand is
\begin{figure}
  \begin{center}
\includegraphics[width=30pc,height=15pc]{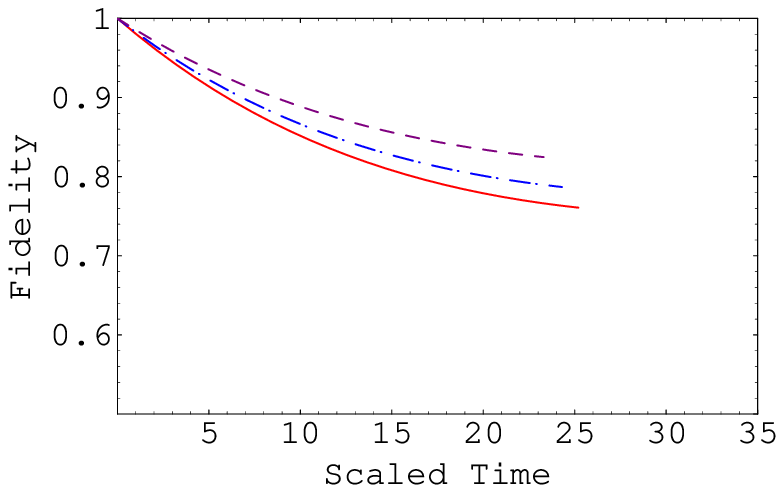}
 \put(-30,160){(a)}\
 \includegraphics[width=30pc,height=15pc]{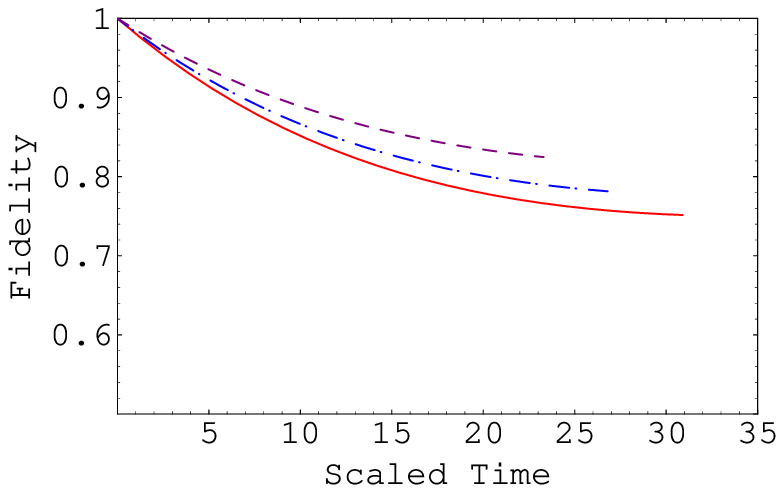}
 \put(-30,160){(b)}
   \caption{The fidelity for (a) Maximally entangled state and (b) Partially entangled stat, where
     $\expect{\sigma_3}_{eq}=1,0.8,0.5. $  for the solid, dash-dot and dot curves respectively,
     with $\alpha_i=0.5$, $\lambda_1=1$
   and $\expect{\tau_3}=-1$.}
    \end{center}
\end{figure}

\begin{equation}
\rho_{Bob}=\eta_1\ket{0}\bra{0}+\eta_2\ket{0}\bra{1}+\eta_3\ket{1}\bra{0}+\eta_4\ket{1}\bra{1},
\end{equation}
where,
\begin{eqnarray}
\eta_1&=&\frac{1}{2}\biggl[|\lambda_1|^2(1+\tilde A_3(t)-\tilde
B_3(t)-\tilde C_{33}(t))+|\lambda_2|^2(1-\tilde A_3(t)+\tilde
B_3(t)+\tilde C_{33}(t))\biggr],
\nonumber\\
\eta_2&=&\frac{1}{2}\left[\lambda_1\lambda^*_2(\tilde
C_{11}(t)-\tilde C_{22}(t))+\lambda_1^*\lambda_2(\tilde
C_{11}(t)+\tilde C_{22}(t))\right],
\nonumber\\
\eta_3&=&\frac{1}{2}\left[\lambda_1\lambda^*_2(\tilde
C_{11}(t)+\tilde C_{22}(t))+\lambda_1^*\lambda_2(\tilde
C_{11}(t)-\tilde C_{22}(t))\right],
\nonumber\\
\eta_4&=&\frac{1}{2}\biggl[|\lambda_1|^2(1+\tilde A_3(t)+\tilde
B_3(t)+\tilde C_{33}(t))+|\lambda_2|^2(1-\tilde A_3(t)+\tilde
B_3(t)-\tilde C_{33}(t))\biggr],
\nonumber\\
\end{eqnarray}
The fidelity, $F$, of the teleported state is
 \begin{equation}
 F=|\lambda_1|^2\eta_1+\lambda_1\lambda_2^*\eta_2+\lambda_1^*\lambda_2\eta_3+|\lambda_2|^2\eta_4.
 \end{equation}

 In Fig$.8$, we plot the fidelity of the teleported state at Bob's hand.
  When the maximally entangled state is used as
a quantum channel between Alice and Bob, the fidelity of the
teleported state is shown in Fig.$8a$. It is clear that,  the
fidelity decreases as one increases the absolute  equilibrium
values. This is due to the decay of the degree of entanglement. On
the other hand, since  the entanglement survives for a long time,
the output state can be used to achieve quantum teleportation for
a long time. Fig.$(8b)$, shows the behavior of the fidelity of the
teleported state when  the output state with $p=0.5$ is used as a
channel. The teleported time, the time which one can use the
output state as channel, increases and the fidelity is better than
that shown in Fig.$(8a)$.

\section{Conclusion}
In this contribution, we  have investigated analytically the
dynamics of a two-qubit state passes through a Bloch channel. We
have discussed  the influence of the Bloch channel's parameters on
the separability and entangled  behavior of the output state. The
results show that, the robustness of the entangled qubit pairs
increases as one increases the absolute values of the equilibrium
parameters or decreases the ratio of the longitudinal and
transverse relaxation times.

The amount of entanglement contained in the output state is
quantified, where we have shown that it is fragile for maximum
entangled states and robust for partial entangled states. The
phenomena of the decay and the sudden death of entanglement are
observed for these types of systems.

Moreover the ability of performing quantum teleportation  by using
the output state is examined. It is found that for  large absolute
equilibrium values, the output state is more useful for quantum
teleportation. Furthermore, the intervals of time in which the
state is available for performing teleportation increase. The
original teleportation protocol is performed by using the output
state as a channel between Alice and Bob. From our results, we
see that the fidelity of the teleported state increases as one
decreases the equilibrium values of the two qubits, but the time
in which the state is useful for quantum teleportation decreases.

\end{document}